 \documentclass[authoryear,preprint,review,12pt]{elsarticle}

\usepackage{newtxtext,newtxmath}
\usepackage[T1]{fontenc}
\usepackage{ae,aecompl}
\usepackage{graphicx}	% Including figure files
\usepackage{amsmath}	% Advanced maths commands

\usepackage{newtxmath}
\usepackage{graphics}
\usepackage[export]{adjustbox}
\usepackage[utf8]{inputenc}
\usepackage[english]{babel}
\usepackage{natbib}
\usepackage{booktabs}
\usepackage{float}
\usepackage{xspace}
\usepackage{hyperref}
\usepackage{tabularx}
\hypersetup{colorlinks,
  citecolor=red,
  linkcolor=red,
  urlcolor=red}
\setcitestyle{authoryear,round}

\usepackage{amssymb}
\usepackage{lipsum}
\journal{Physica A}

\begin{document}

\begin{frontmatter}

%% Title, authors and addresses

%% use the tnoteref command within \title for footnotes;
%% use the tnotetext command for theassociated footnote;
%% use the fnref command within \author or \affiliation for footnotes;
%% use the fntext command for theassociated footnote;
%% use the corref command within \author for corresponding author footnotes;
%% use the cortext command for theassociated footnote;
%% use the ead command for the email address,
%% and the form \ead[url] for the home page:
%% \title{Title\tnoteref{label1}}
%% \tnotetext[label1]{}
%% \author{Name\corref{cor1}\fnref{label2}}
%% \ead{email address}
%% \ead[url]{home page}
%% \fntext[label2]{}
%% \cortext[cor1]{}
%% \affiliation{organization={},
%%            addressline={}, 
%%            city={},
%%            postcode={}, 
%%            state={},
%%            country={}}
%% \fntext[label3]{}

\title{Anomalous Entropic Behavior Observed in Quasar 3C 273}

%% use optional labels to link authors explicitly to addresses:
%% \author[label1,label2]{}
%% \affiliation[label1]{organization={},
%%             addressline={},
%%             city={},
%%             postcode={},
%%             state={},
%%             country={}}
%%
%% \affiliation[label2]{organization={},
%%             addressline={},
%%             city={},
%%             postcode={},
%%             state={},
%%             country={}}

\author[1]{C. V. da Silva}

\author[1]{F. V. Alencar Filho}

\author[2]{J. P. Bravo}

\author[1]{D. B. de Freitas}

\affiliation[1]{Departamento de Fisica, Universidade Federal do Ceara, Caixa Postal 6030, Campus do Pici, 60455-900 Fortaleza, Ceara, Brazil}

\affiliation[2]{Avignon Universite, Campus Jean-Henri Fabre Centre d'Enseignement et de Recherche en Informatique Agroparc BP 91228, 84911 Avignon Cedex 9, France}

\begin{abstract}
%% Text of abstract
We investigate the flux intensities spanning from radio waves to $\gamma$-rays across 36 light curves of Quasar 3C 273, utilizing publicly available data collected by the Integral Science Data Centre (ISDC) database. Our analysis reveals a consistent adherence of all light curves from this quasar to $q$-Gaussian distribution. This compelling finding strongly suggests a nonextensive behavior exhibited by Quasar 3C 273. Moreover, we derive the $q$ entropic indices for these light curves, providing insights into the degree of nonextensivity, where cases with $q>1$ were primarily observed. Utilizing this index, we estimate the nonextensive entropy ($S_{q}$) and explore its correlation with the energy (in eV) and the $q$ index. Notably, we observe a tendency for the $q$ value to increase as the Tsallis entropy decreases. Remarkably, our most significant observation pertains to the relationship between the entropy $S_{q}$ and the energy of the source. We identify an anomalous behavior in entropy, particularly evident in the infrared and $\gamma$-ray wavebands.
\end{abstract}

%%Graphical abstract
%\begin{graphicalabstract}
%\includegraphics{grabs}
%\end{graphicalabstract}

%%Research highlights
%\begin{highlights}
%\item Research highlight 1
%\item Research highlight 2
%\end{highlights}

\begin{keyword}
%% keywords here, in the form: keyword \sep keyword, up to a maximum of 6 keywords
Quasar 3C 273 \sep Nonextensive statistical mechanics \sep Active Galactic Nuclei (AGNs) \sep Light curves

%% PACS codes here, in the form: \PACS code \sep code

%% MSC codes here, in the form: \MSC code \sep code
%% or \MSC[2008] code \sep code (2000 is the default)

\end{keyword}

\end{frontmatter}

%\tableofcontents

%% \linenumbers

%% main text

\section{Introduction}
Active Galactic Nuclei (AGNs) are known for their extreme brightness and high variability across multiple wavelengths \citep{1979ApJ...232...34B,1995PASP..107..803U,galaxies11050096}. The Quasar 3C 273, a bright and nearby radio-loud quasar, has been extensively studied for its flux curves at all frequencies over several decades, making it one of the most well-documented AGNs \citep{1963Natur..197,1987A&A...176..197C,1990A&A...234...73C,2008MmSAI..79.1011C}. The Integral Science Data Centre (ISDC) database and the High-Energy Astrophysics Virtually ENlighted Sky (HEAVENS) web interface have provided a wealth of observational data, including light curves covering more than 40 years of observations across various wavebands \citep{1999A&AS..134...89T,2008A&A...486..411S,2010int..workE.162W}. In these studies, the variability of Quasar 3C 273 at high energies, from keV to GeV, has been analyzed using observations from instruments such as IBIS, SPI, JEM-X on board INTEGRAL, PCA on board RXTE, and LAT on board Fermi \citep{1996SPIE.2808...59J,2009ApJ...697.1071A}.

The variability of AGNs, including Quasar 3C 273, has been studied using different procedures to quantify amplitude and periodicity, revealing complex emission behavior likely originating from different locations within the AGN \citep{2008A&A...486..411S}. Various methods, such as the discrete autocorrelation function, the Lomb-Scargle periodogram, and the wavelet transform, have been employed to study the quasi-periodic and periodic variability of AGN sources and their correlations with several statistical parameters, such as standard deviation of signal, thermodynamical properties and short- and long-term fluctuations \citep{1998ApJ...503..662H,2006A&A...456L...1K,2007A&A...469..899H}. Additionally, based on several studies, e.g., \cite{2006AdSpR..38.1405V,2016MNRAS.461.3145V,2015Natur.518...74G}, reports of statistical analysis of Quasar 3C 273 have provided relevant insights into long- and short-scale variations with a higher number of good data points from radio to $\gamma$--ray wavebands.

The literature still lacks a significant range of papers that treat quasars as out-of-thermodynamic equilibrium energy sources. An important work in this research line is \cite{ROSA20136079}'s study, which emphasizes the role of non-extensive statistical mechanics in investigating the entropic behavior of various astrophysical sources from Novae to Pulsars. However, no studies have yet analyzed the entropic behavior of quasars across different wavelengths. In this context, non-equilibrium statistics, particularly non-extensive statistical mechanics, are crucial for understanding how self-gravitating systems manage energy retention and release mechanisms \citep{tsallis1988,Tsallis4}. 

In the astrophysical context, nonextensive entropy, such as Tsallis entropy, has been suggested for generalized entropy of self-gravitating systems and has been extensively examined from astrophysical viewpoints \citep{1995PASP..107..803U,1979ApJ...232...34B,2008A&A...486..411S}. In the context of galaxies, \cite{NAKAMICHI2002595} showed that nonextensive statistics applied to the galaxy distributions and the overall structure revealed a ubiquitous instability of self-gravitating systems since they obtained the negative parameter $q$. Furthermore, in active galactic nuclei (AGN) and quasars, nonextensive statistical mechanics has provided insights into energetic particle distributions, radiation mechanisms, and the properties of these objects \citep{10.1140/epja/i2011-11052-1}. The application of nonextensive statistics to gravitational waves has also opened new avenues for estimating parameters of binary systems around supermassive black holes and analyzing the thermodynamic properties of gravitational wave sources \citep{2023EL....14159002D}. These studies highlight the versatility and significance of nonextensive statistical mechanics in unraveling the complexities of astrophysical systems, offering valuable insights into galaxies, AGN, and quasars.

Our paper analyzes the Quasar 3C 273 from radio waves to $\gamma$-ray as a nonextensive phenomenon. In addition, we study their degree of nonextensivity, analyzing the distributions of 36 light curves. Our main aim is to investigate the entropic behavior as a function of the energy of the source. This paper is organized as follows: Section 2 describes the data and procedures used. In Sect. 3, we report the results and discussions, including a comparison with the literature. Conclusions and future works are presented in the last Section.

\section{Working sample}\label{data}

\begin{table}
\scriptsize
\caption[]{\label{tabobs} Observational data selection from the datasets hosted in the ISDC database of Quasar 3C 273.}
\begin{center}
\begin{tabular}{cccc}
 \hline
 \noalign{\smallskip}
 \multicolumn{4}{c}{{\bf ISDC Database}}\\
 \noalign{\smallskip}
 \hline
\noalign{\smallskip}
Spectral Emission & Waveband & Date Range & $\#$ Obs.\\
\noalign{\smallskip}
\hline
\noalign{\smallskip}	
Radio					 & 5 GHz & 1967-2006 & 870\\
 & 8 GHz & 1963-2006 & 1568\\
												 & 15 GHz & 1963-2006 & 1286\\
												 & 22 GHz & 1976-2004 & 1099\\
												 & 37 GHz & 1970-2006 & 1259\\
\hline												
\noalign{\smallskip}
Millimeter				 & 3.3 mm & 1965-2006 & 1548\\
and submillimeter 			 & 2.0 mm & 1981-1997 & 203\\
												 & 1.3 mm & 1981-2007 & 545\\
             & 1.1 mm & 1973-2007 & 322\\
											     & 0.8 mm & 1981-2007 & 496\\										
\hline
\noalign{\smallskip}
Infrared						 & L(3.6 $\mu$m) & 1969-1997 & 164\\
												 & K(2.2 $\mu$m) & 1967-2004 & 350\\
												 & H(1.65 $\mu$m) & 1967-2004 & 317\\
												 & J(1.25 $\mu$m) & 1976-2004 & 280\\
\hline
\noalign{\smallskip}
Optical 						 & R(7000 \AA) & 1977-2002 & 185\\
& G(5798 \AA) & 1985-2003 & 438\\
												 & V(5479 \AA) & 1968-2005 & 730\\
                                                 & V1(5395 \AA) & 1985-2003 & 438\\
                                                 & B2(4466 \AA) & 1985-2003 & 438\\
												 & B(4213 \AA) & 1968-2005 & 755\\
                                                & B1(4003 \AA) & 1985-2003 & 438\\
												 & U(3439 \AA) & 1968-2005 & 680\\
\hline
\noalign{\smallskip}
Ultraviolet 					 & 3000 \AA & 1978-2005 & 210\\
& 2700 \AA & 1978-1996 & 199\\
& 2425 \AA & 1978-1996 & 210\\
& 2100 \AA & 1978-2005 & 191\\
												 & 1950 \AA & 1978-1996 & 237\\
												 & 1700 \AA & 1978-1996 & 238\\
             & 1525 \AA & 1978-1996 & 239\\
												 & 1300 \AA & 1978-1996 & 235\\												
\hline																																								\noalign{\smallskip}																		
X-ray and $\gamma$-ray 							 & 5 KeV & 1970-2005 & 1032\\
												 & 10 KeV & 1974-2005 & 1026\\
             & 20 KeV & 1976-2005 & 1093\\
												 & 50 KeV & 1977-2005 & 1108\\
												 & 100 KeV & 1978-2005 & 1107\\
             &200 KeV & 1978-2005 & 149\\
\hline																				
																
\end{tabular}
\end{center}
\end{table}

\begin{figure}
\centering
\includegraphics[scale=0.5]{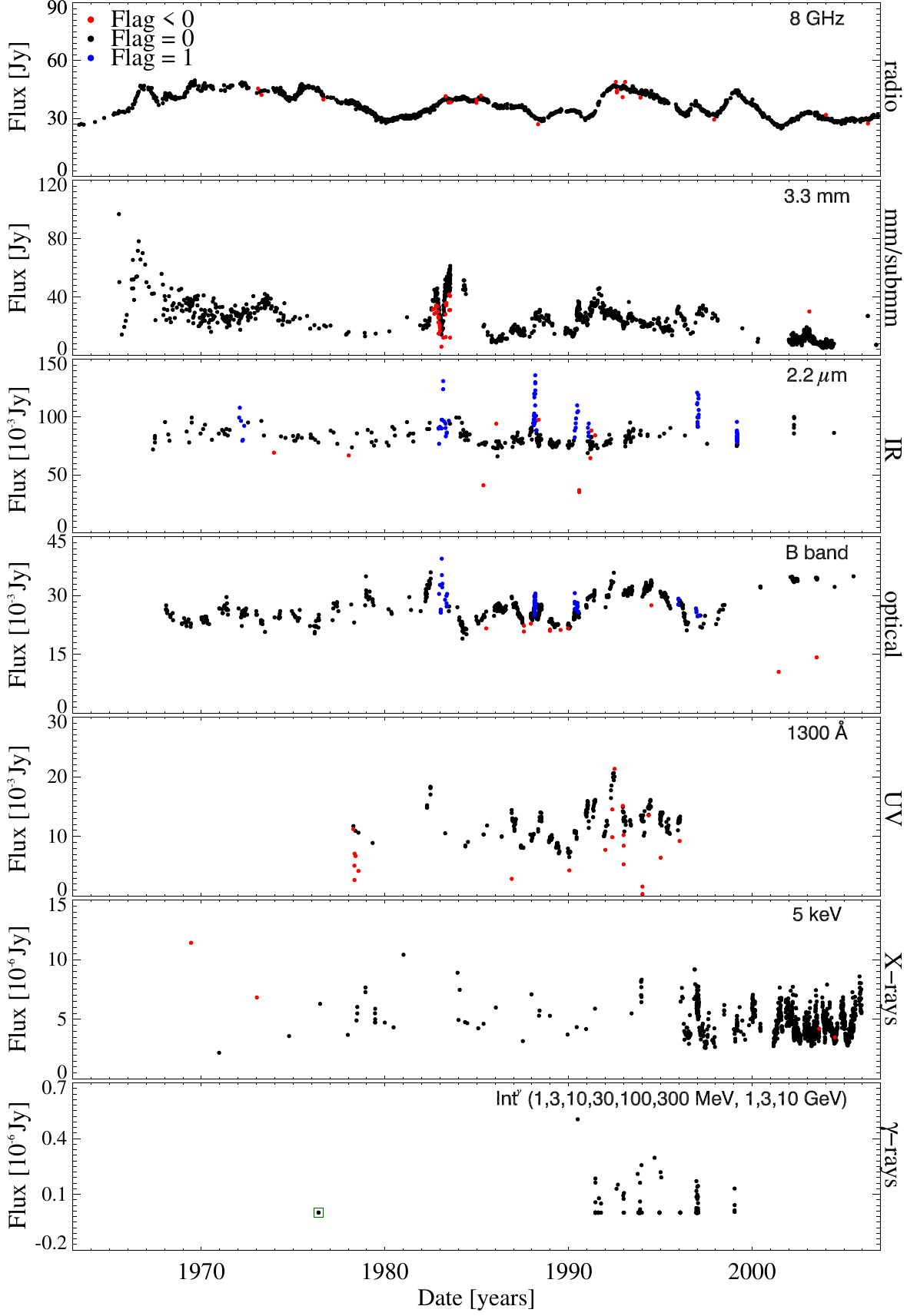}
\caption{Examples of light curves of the quasar 3C 273 used in our analysis at different wavelengths: 8 GHz, 3.3 mm, 2.2 $\mu$m (K band), 4213 \AA (B band), 1300 \AA, 5 keV. The synchrotron flares (\texttt{Flag}=1) are highlighted in blue for the infrared and optical datasets. The remaining good data (\texttt{Flag}=0) are shown in black, and the data disregarded (\texttt{Flag}<0) in red. The green square in $\gamma$-rays points to the observations removed, preventing big gaps in the signal. The light curves from 1 MeV to 10 GeV (referred to as Int$^\gamma$ for $\gamma$-rays) extracted from he HEAVENS interface are solely included to highlight the limited amount of data available for analysis.}
\label{FigLCs}
\end{figure}

Even though the ISDC 3C 273's database gives access to many multi-wavelength light curves from radio to $\gamma$-ray ranges \citep{2008A&A...486..411S,1999A&AS..134...89T}, we choose those having sufficient data taken over a long time span since our goal is to study the long-term variability of this AGN. Therefore, from radio waves to $\gamma$-rays, several wavelengths are selected for each spectral emission. Thus, we have selected 36 flux curves among the 70-time series the ISDC database provides. Each observation is related to a flag parameter, allowing us to determine our sample using only good and reliable data. Good data have a \texttt{Flag} of 0 or 1, where 1 is used for optical and infrared observations with significant contributions from synchrotron flares. ``Useless'', ``uncertain'' or ``dubious'' data points are indicated by a negative \texttt{Flag}, which we disregard. Our sample is exposed in Table \ref{tabobs}. The date range (between the first and the last observation) and the number of data points correspond to reliable data (\texttt{Flag}$\ge$0). An example of each spectral band light curve is shown in Fig. \ref{FigLCs}. Due to the insufficient data from the RXTE-PCA and Fermi-LAT retrieved from the HEAVENS interface, we decided to discard these data from our sample and only considered data from the ISDC database. However, we have included in the bottom panel of Figure \ref{FigLCs}, the integrated Int$^\gamma$ data distributed in the wide energy range that goes from 1 MeV to 10 GeV, where it is possible to verify the scarcity of data visually.

\begin{figure}
\centering
 \includegraphics[scale=0.5]{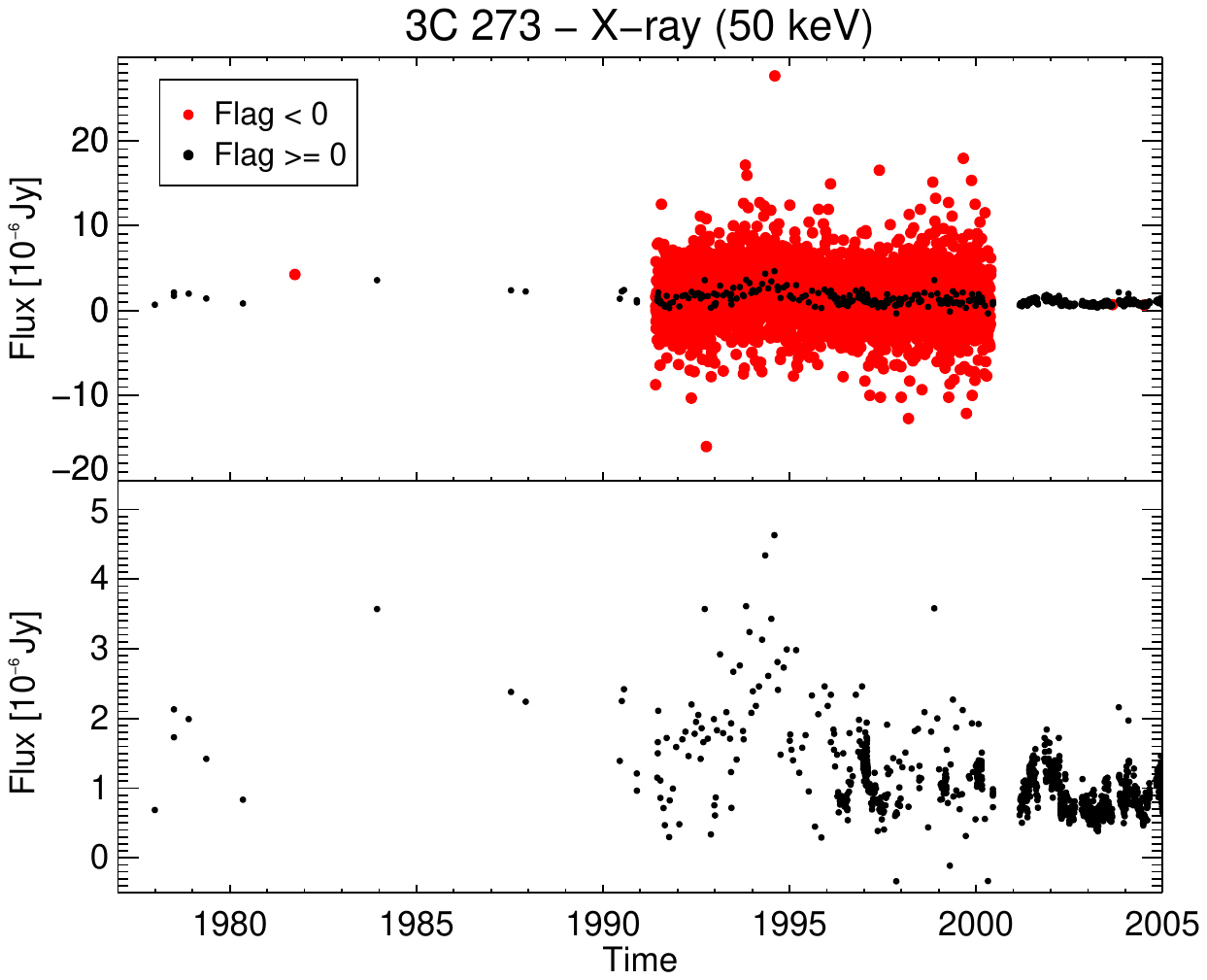}
\caption{Light curve of the quasar 3C 273 at 50 keV covering 28 years. Top panel: The full light curve with less reliable data is shown in red. Bottom panel: good data (in black) remaining after removal of the observational points linked to the negative Flag parameter.}
\label{FigLC50kev}
\end{figure}

Covering up to 43 years, radio light curves are featured to present profiles of intense variability and periodicity, as well as millimeter and submillimetre (hereafter mm/sub-mm) flux curves, while those in infrared (IR) are distinguished for their low variability. No treatment of time series selected is performed, just some points are disregarded since big gaps in the observations could highly contribute in the analysis and distort the components in the whole signal, causing a misinterpretation of the variability. Considering this, the time span of the flux curve in the infrared L band (3.6 $\mu$m) is reduced from 35 to 28 years since there is a gap from 1997 to 2004; in fact, one single data point at this date interval is removed. The same is done at 2.0 mm, decreasing the time span from 23 to 16 years. In our X-ray emission sample, many less reliable data points are observed at 50, 100, and 200 keV, as shown in red in the top panel of Fig. \ref{FigLC50kev}. However, more refined light curves are obtained after removing the observations with a flag parameter of less than 0, with the remaining ones with \texttt{Flag}=0 (bottom panel). 

These meticulous adjustments and considerations in the data treatment ensure the reliability and robustness of the dataset used for the analysis, thereby enhancing the validity and accuracy of the study's findings.

\begin{figure}
\centering
 \includegraphics[scale=0.5]{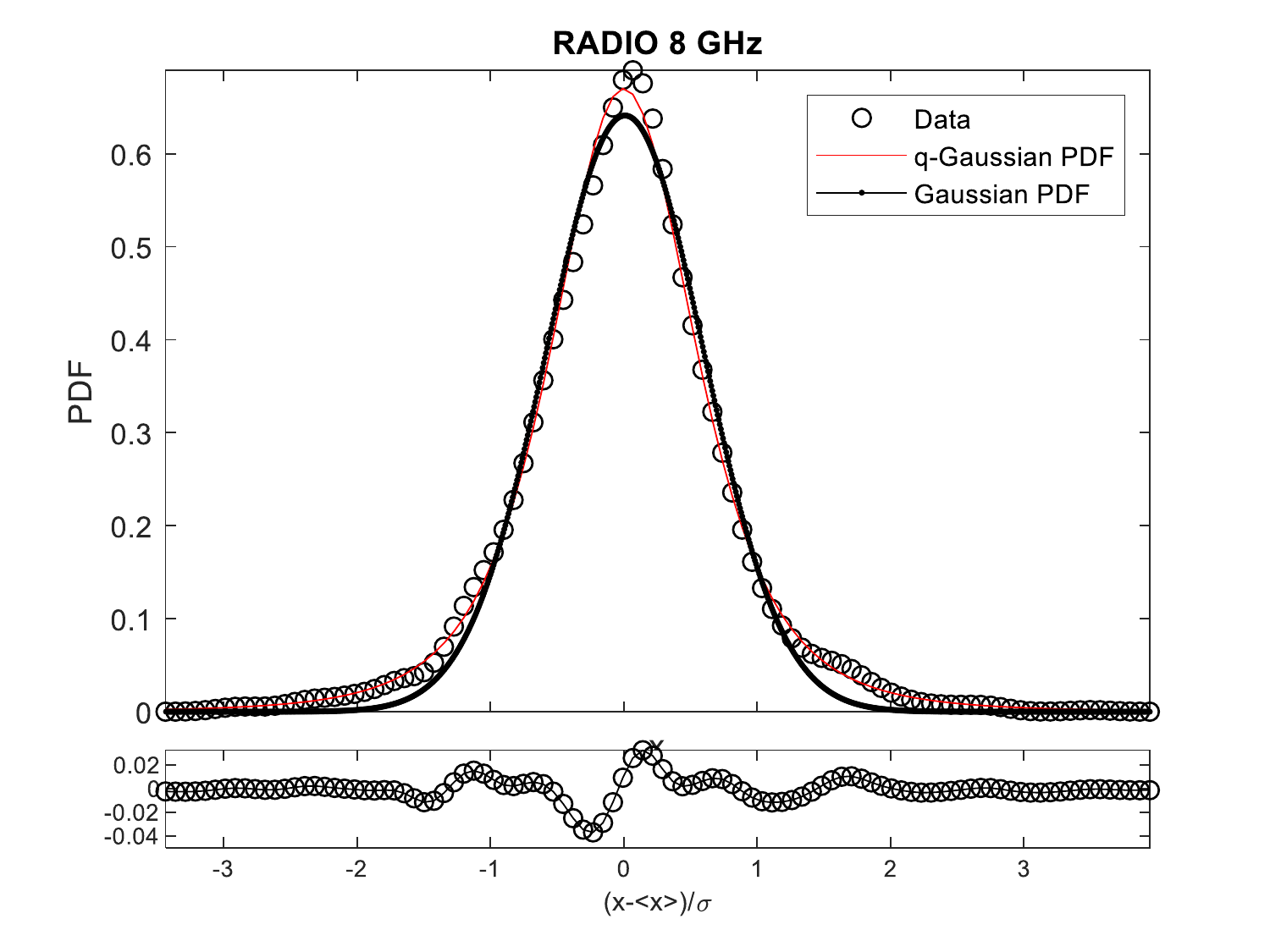}
\caption{PDFs of flux for Radio 8 GHz light curve. The open circles represent data; the red and black curves fit with a $q$-Gaussian (Tsallis) distribution and standard Gaussian, respectively. The bottom panel denotes the residual between the empirical distribution and the $q$-Gaussian PDF.}
\label{FigqRadio8GHz}
\end{figure}

\begin{figure}
\centering
 \includegraphics[scale=0.5]{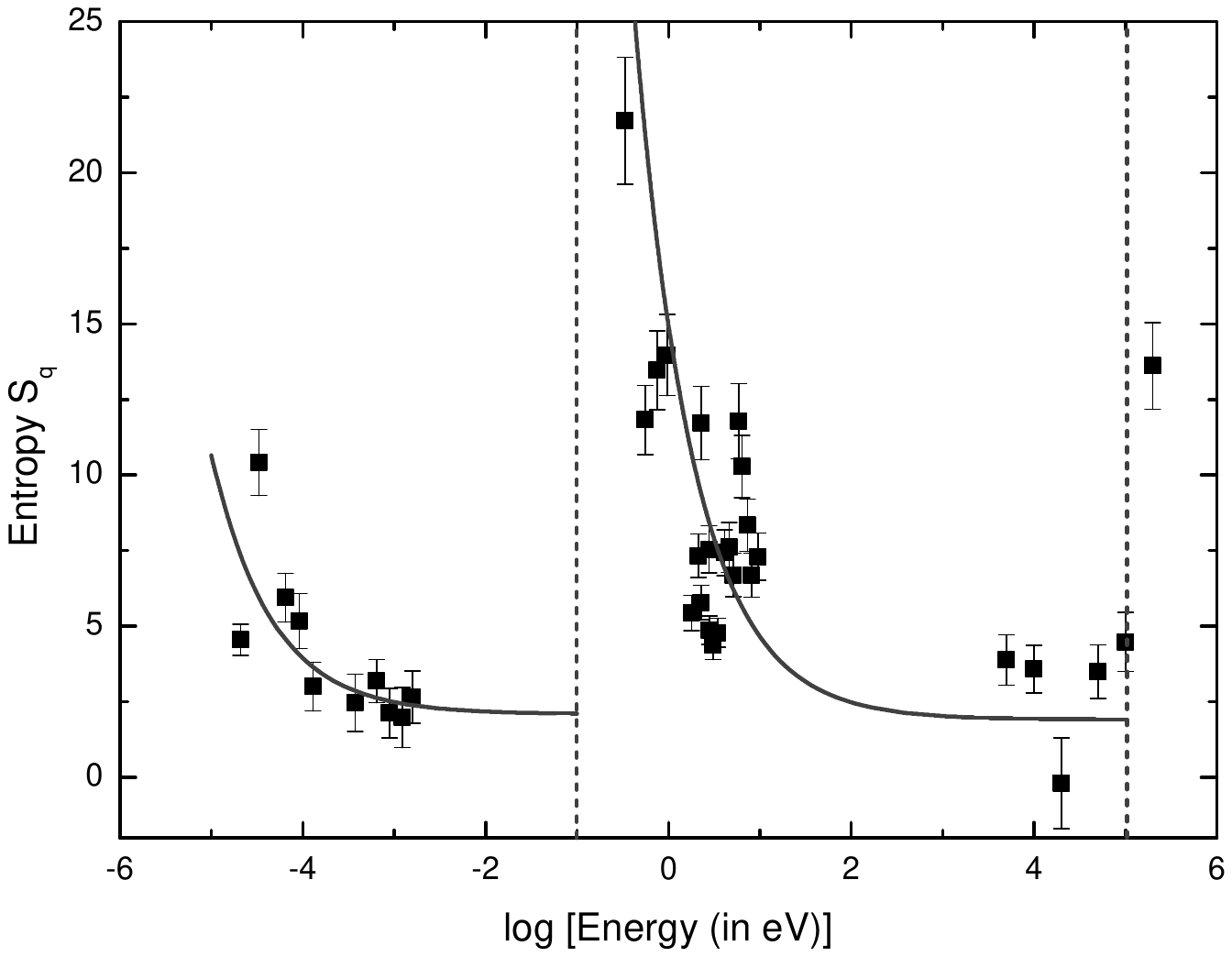}
\caption{Tsallis entropy values $S_{q}$ as a function of the logarithm of energy band (in eV) for all 36 light curves of Quasar 3C 273. The gray curves denote the best adjustment using the non-linear L-M method. The vertical dashed lines denote the asymptotic behavior of nonextensive entropy in a specific energy range.}
\label{Fig2}
\end{figure}

\begin{figure}
\centering
 \includegraphics[scale=0.5]{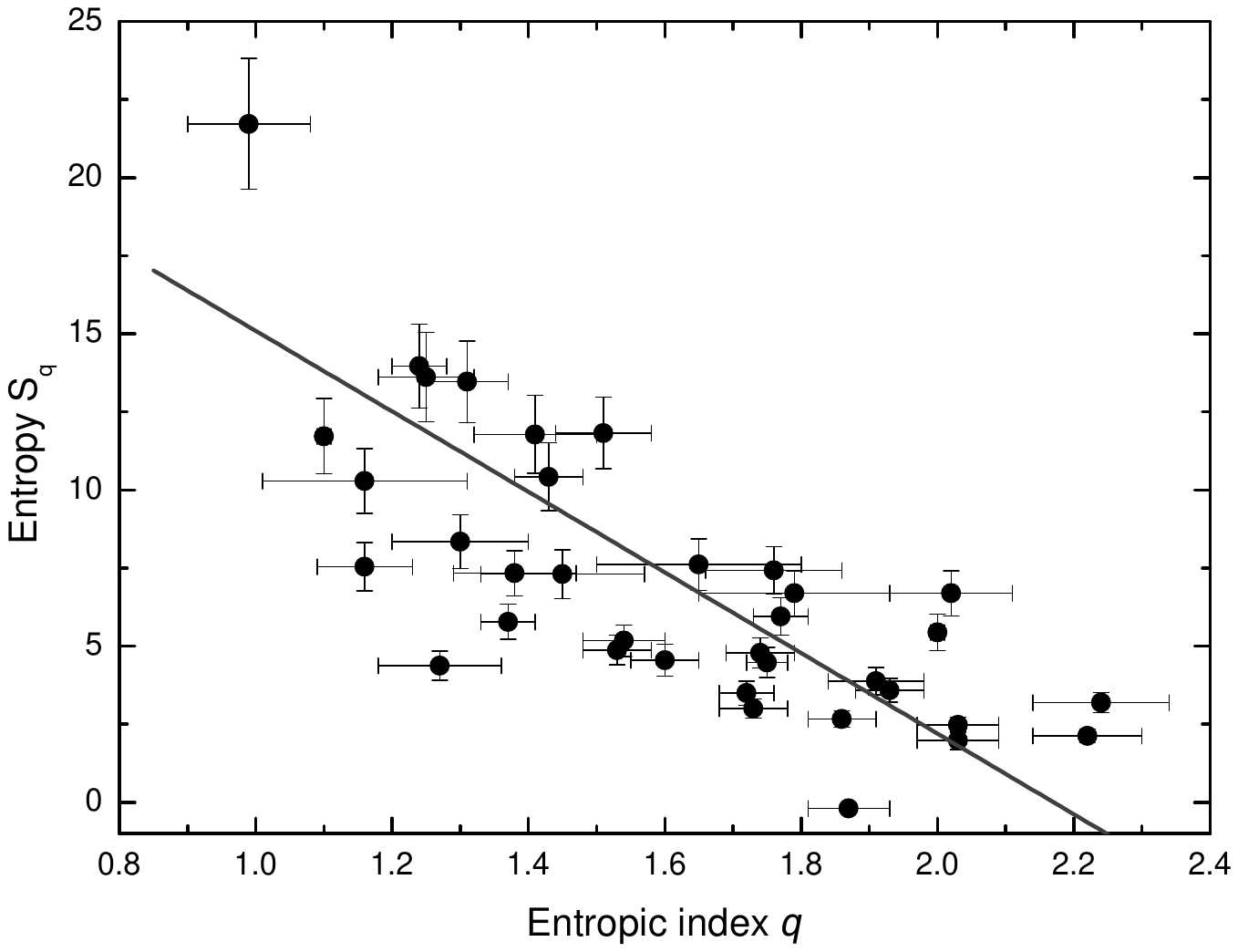}
\caption{Tsallis entropy values $S_{q}$ as a function of the $q$-entropic index for all 36 light curves of Quasar 3C 273.}
\label{Fig3}
\end{figure}

\begin{table}
\scriptsize
\caption[]{\label{tabresults} Best $q$, $\sigma_{q}$, $S_{q}$ values and their confidential intervals determined using non-linear regression L–M method for all the wavebands.}
\begin{center}
\begin{tabular}{c|c|c|c|c}
 \hline
 \noalign{\smallskip}
 \multicolumn{5}{c}{{\bf ISDC Database}}\\
 \noalign{\smallskip}
 \hline
\noalign{\smallskip}
Spectral Emission & Energy (eV) & $q$ & $\sigma_{q}$ & $S_{q}$\\
\noalign{\smallskip}
\hline
\noalign{\smallskip}	
Radio 						 & $2.07\times 10^{-5}$ & $1.60^{1.65}_{1.54}$ &         
                                                $0.34^{0.35}_{0.33}$ & $4.55^{5.09}_{4.07}$\\
 & $3.31\times 10^{-5}$ & $1.43^{1.48}_{1.39}$ &         
                                                $0.80^{0.82}_{0.78}$ & $10.42^{11.56}_{9.38}$\\
												 & $6.41\times 10^{-5}$ & $1.77^{1.81}_{1.73}$ & $0.74^{0.76}_{0.72}$ & $5.95^{6.75}_{5.15}$\\
												 & $9.16\times 10^{-5}$ & $1.54^{1.60}_{1.48}$ & $1.19^{1.23}_{1.16}$ & $5.17^{6.08}_{4.27}$\\
												 & $1.30\times 10^{-4}$ & $1.73^{1.78}_{1.68}$ & $0.89^{0.92}_{0.86}$ & $3.00^{3.81}_{2.21}$\\
\hline												
\noalign{\smallskip}
Millimeter 				 & $3.72\times 10^{-4}$ & $2.03^{2.09}_{1.98}$ & $1.45^{1.50}_{1.40}$ & $2.47^{3.42}_{1.54}$\\
and submillimeter 			 & $6.53\times 10^{-4}$ & $2.24^{2.34}_{2.13}$ & $1.04^{1.12}_{0.96}$ & $3.19^{3.89}_{2.48}$\\
												 & $8.85\times 10^{-4}$ & $2.22^{2.30}_{2.14}$ & $0.69^{0.73}_{0.65}$ & $2.12^{2.94}_{1.31}$\\
            & $1.24\times 10^{-3}$ & $2.03^{2.09}_{1.97}$ & $0.25^{0.26}_{0.24}$ & $1.98^{2.98}_{1.01}$\\
											     & $1.57\times 10^{-3}$ & $1.86^{1.91}_{1.81}$ & $0.74^{0.76}_{0.72}$ & $2.66^{3.52}_{1.83}$\\										
\hline
\noalign{\smallskip}
Infrared 						 & $0.34$ & $0.99^{1.08}_{0.90}$ & $1.08^{1.12}_{1.04}$ & $21.72^{23.91}_{19.72}$\\
												 & 0.56 & $1.51^{1.58}_{1.44}$ & $0.65^{0.67}_{0.62}$ & $11.82^{13.02}_{10.74}$\\
												 & 0.75 & $1.31^{1.37}_{1.24}$ & $0.77^{0.79}_{0.75}$ & $13.46^{14.83}_{12.22}$\\
												 & 0.98 & $1.24^{1.28}_{1.21}$ & $0.83^{0.84}_{0.81}$ & $13.97^{15.38}_{12.69}$\\
\hline
\noalign{\smallskip}
Optical 						 & 1.80 & $2.00^{2.01}_{1.94}$ & $0.28^{0.29}_{0.27}$ & $5.44^{6.06}_{4.89}$\\
& 2.14 & $1.38^{1.47}_{1.29}$ & $0.55^{0.57}_{0.52}$ & $7.33^{8.09}_{6.65}$\\
												 & 2.30 & $1.37^{1.41}_{1.32}$ & $0.56^{0.57}_{0.55}$ & $5.78^{6.37}_{5.24}$\\
              & 2.31 & $1.10^{1.11}_{1.09}$ & $2.47^{2.48}_{2.46}$ & $11.72^{12.97}_{10.55}$\\
              & 2.78 & $1.16^{1.23}_{1.10}$ & $0.60^{0.62}_{0.58}$ & $7.54^{8.36}_{6.80}$\\
												 & 2.82 & $1.53^{1.58}_{1.48}$ & $0.45^{0.46}_{0.43}$ & $4.87^{5.37}_{4.42}$\\
             & 3.10 & $1.27^{1.36}_{1.18}$ & $0.43^{0.45}_{0.41}$ & $4.37^{4.85}_{3.92}$\\
												 & 3.44 & $1.74^{1.79}_{1.70}$ & $0.42^{0.43}_{0.41}$ & $4.78^{5.28}_{4.33}$\\
\hline
\noalign{\smallskip}
Ultraviolet					 & 4.13 & $1.76^{1.86}_{1.65}$ & $0.42^{0.45}_{0.40}$ & $7.43^{8.23}_{6.71}$\\
& 4.59 & $1.65^{1.80}_{1.50}$ & $0.47^{0.50}_{0.43}$ & $7.61^{8.47}_{6.82}$\\
& 5.11 & $2.02^{2.11}_{1.93}$ & $0.32^{0.33}_{0.30}$ & $6.69^{7.44}_{5.99}$\\
& 5.91 & $1.41^{1.50}_{1.32}$ & $0.72^{0.76}_{0.69}$ & $11.78^{13.08}_{10.58}$\\
												 & 6.36 & $1.16^{1.31}_{1.02}$ & $0.69^{0.74}_{0.65}$ & $10.28^{11.36}_{9.29}$\\
												 & 7.29 & $1.30^{1.40}_{1.19}$ & $0.59^{0.62}_{0.56}$ & $8.34^{9.24}_{7.51}$\\
& 8.13 & $1.79^{1.93}_{1.65}$ & $0.41^{0.45}_{0.38}$ & $6.69^{7.46}_{6.01}$\\
												 & 9.54 & $1.45^{1.57}_{1.32}$ & $0.55^{0.59}_{0.52}$ & $7.30^{8.13}_{6.55}$\\												
\hline																																								\noalign{\smallskip}																		
X-ray and $\gamma$-ray 							 & $5\times 10^{3}$ & $1.91^{1.98}_{1.85}$ & $0.44^{0.45}_{0.42}$ & $3.88^{4.71}_{3.07}$\\
												 & $ 10^{4}$ & $1.93^{1.98}_{1.87}$ & $0.43^{0.45}_{0.42}$ & $3.58^{4.34}_{2.82}$\\
             & $2\times 10^{4}$ & $1.87^{1.93}_{1.82}$ & $0.17^{0.18}_{0.16}$ & $-0.19^{1.31}_{-1.29}$\\
												 & $5\times 10^{4}$ & $1.72^{1.76}_{1.68}$ & $0.33^{0.34}_{0.32}$ & $3.49^{4.34}_{2.65}$\\
												 & $ 10^{5}$ & $1.75^{1.78}_{1.72}$ & $0.48^{0.49}_{0.47}$ & $4.48^{5.46}_{3.52}$\\	
             & $2\times 10^{5}$ & $1.25^{1.32}_{1.17}$ & $1.19^{1.24}_{1.15}$ & $13.61^{15.24}_{12.37}$\\
\hline																																								\noalign{\smallskip}																												
\end{tabular}
\end{center}
\end{table}

\section{Nonextensive theoretical framework}
A compelling approach to scrutinizing the entropic dynamics of Quasar 3C 273 involves examining how nonextensive entropy, segregated by wavelength, varies with energy. In the literature, explanations for the intricate behavior of this phenomenon often rely on conventional Boltzmannian exponentials \citep[cf.][]{defreitasetal15, defreitas2020, defreitas2021, Burlaga2004-lt, Leubner2005one, Leubner2005two, De_Freitas2009-fz, Du2010-qv, Rouillard2021-cd}. Characterizing the intricate flux behavior, particularly concerning the distribution across different wavebands, is pivotal for unraveling the physical underpinnings of the distribution tail. Such distributions are inherently rooted in entropy. Among the various out-of-equilibrium Boltzmann-Gibbs (B-G) entropies found in the literature are those by \cite{druy1930,druy1934}, \cite{ren}, \cite{sharma}, \cite{abe1997}, \cite{papa}, \cite{borges1}, \cite{lands}, \cite{anteneodo1999}, \cite{frank}, \cite{kani}, and finally, the Tsallis nonextensive entropy.

A significant advantage of nonextensive statistical mechanics lies in its capacity to elucidate a broad spectrum of systems exhibiting out-of-equilibrium thermodynamic behavior \citep{Tsallis2003-tq}. For instance, systems featuring long-range interactions, such as X-rays emitted from astrophysical sources \citep{ROSA20136079}, are exemplary cases. These systems manifest in various astrophysical phenomena such as cataclysmic variables, X-ray binary systems, pulsars, and quasars \citep{PhysRevE.68.041104, RePE}. Consequently, nonextensive statistical mechanics offers a framework to map radio flux variations up to frequencies in the hundreds of gigahertz range, thereby tracing energy distribution in distant sources via their spatial disposition \citep{10.3390/atoms7010018}.

Conversely, a potential drawback of the nonextensive approach lies in the quantity of available sample data. Understanding this aspect necessitates consideration of the Central Limit Theorem (CLT) \citep{Umarov2008-yg}. According to the CLT, the sampling distribution of the mean of a large number of independent, identically distributed random variables converges toward a Gaussian distribution, regardless of the underlying population distribution \citep{Kwak2017-fi}. As the sample size increases, the sampling distribution of the mean tends more towards Gaussian, reducing the standard error of the mean and, consequently, more precise estimates of the population parameters \citep{Kwak2017-fi}. Conversely, as the sample size diminishes, the sampling distribution becomes increasingly skewed and less Gaussian, resulting in less accurate estimations.

In Tsallis' $q$-entropy, the classical statistical formulation is not followed, and entropy additivity does not occur. The new entropic approach is given by $S(A+B)=S(A)+S(B)+(1-q)S(A)S(B)$, where $S$ denotes the entropy. The last term of this equation exhibits the interaction between systems $A$ and $B$ that does not exist in the extensive formalism \citep{tsallis1988,Tsallis3,Tsallis4}. At the core of Tsallis' nonextensive statistical mechanics is the Tsallis entropy, which is a generalization of the B-G entropy. Thus, the Tsallis entropy is given by:
\begin{equation}\label{g4}
        S_{q}=k\left(\frac{1-\sum^{W}_{n=1}p^{q}_{n}}{q-1}\right) \quad (q \in\mathbb{R}),
\end{equation}
where $q$ is the entropic index that characterizes the generalization, $k$ is Boltzmann's constant, and $p_{n}$ 
are the probabilities associated with $W$ (microstates in the system) configurations ($W\in \mathbb{N}$) with the condition $\sum_{n} p_{n}=1$. The extra term denotes the interaction between systems A and B. For $q\rightarrow1$ and $p_{n}=1/W$, B-G entropy, defined as $S_{B-G}=k\ln W$, is recovered. 

In the present study, the suitable $q$-values are obtained from the probability distribution function (PDF) associated with fluctuations of Quasar 3C 273 flux (in Jy). The PDF distributions defined as $q$-Gaussians can be generated by the Tsallis generalized entropy $S_{q}$ maximization process. In particular, the PDF curves were always normalized by subtracting from the $x$’s their average $\langle x\rangle$ and dividing by the correspondent standard deviation $\sigma$, i.e., $y=(x-\langle x\rangle)/\sigma$, following the same methodology from \cite{10.1117/12.772041}. In this way, we obtain the PDFs given by
    \begin{equation}\label{g12}
        \mathrm{PDF}=A_{q}\left[1+(q-1)\left(\frac{y}{\sigma_{q}}\right)^{2}\right]^{1/(1-q)}, \quad q<3,
    \end{equation}
where the normalization constant $A_{q}$ is obtained for two distinct intervals of $q$,
    \begin{equation}
        A_{q}=\frac{\Gamma\left(\frac{5-3q}{2-2q}\right)}{\Gamma\left(\frac{2-q}{1-q}\right)}\sqrt{\left(\frac{1-q}{\pi}\right)\beta_{q}}, \quad q<1,
    \end{equation}
and,
    \begin{equation}\label{g14}
        A_{q}=\frac{\Gamma\left(\frac{1}{q-1}\right)}{\Gamma\left(\frac{3-q}{2q-2}\right)}\sqrt{\left(\frac{q-1}{\pi}\right)\beta_{q}}, \quad q>1,
    \end{equation}
    where
    \begin{equation}\label{g15}
        \beta_{q}=\left[(3-q)\sigma^{2}_{q}\right]^{-1}
    \end{equation}
    and
    \begin{equation}\label{g16}
        \sigma^{2}_{q}=\sigma^{2}\left(\frac{5-3q}{3-q}\right),
    \end{equation}
where $\sigma_{q}$ means the generalized standard deviation as a function of $q$, whereas that $\sigma$ is the canonical standard deviation could be said earlier.

\section{Results and Discussions}\label{results}

It's noteworthy that the majority of observed energy bands exhibit a notable adherence to Tsallis distributions. This tendency might stem from the fact that distributions that more accurately depict the intensity fluctuations of Quasar 3C 273 light curves tend to have elongated tails (as indicated by the values of $q$ highlighted in Table \ref{tabresults}, representing the degree of nonextensivity).

In light of equation \ref{g12}, the proposed solution precisely aligns with the generalization of Gaussian curves, a method widely employed across various astrophysical contexts \citep[cf.][]{De_Freitas2009-fz,defreitas2020}. Notably, the $q$-Gaussian fitting (refer to Fig. \ref{FigqRadio8GHz}) emerges as the optimal choice for modeling Quasar 3C 273 light curves, as corroborated in Table \ref{tabresults}. Specifically, for the 8 GHz Radio light curve, the $q$-Gaussian fitting demonstrates a correlation coefficient of $R^{2}= 0.997$ (indicated by the red line) obtained by a nonlinear regression method based on the Levenberg-Marquardt (L-M) algorithm \citep[cf.][]{83b09f23-b20e-3617-8f72-24765b713f7b,doi:10.1137/0111030}. Conversely, the Gaussian fitting (denoted by the gray dashed line) fails to capture the observed 8 GHz Radio values beyond 1 standard deviation. This pattern persists across all light curves, except within the infrared band at 0.34 eV, where a Gaussian profile prevails, accompanied by the lowest $q$-index value in our dataset. Notably, the selection of the 8 GHz Radio light curve was arbitrary.

%%%%%%%%%%%%%%%%%%
Figure \ref{Fig2} reveals an anomalous relationship between the non-extensive entropy and the energy levels of Quasar 3C 273 across various wavelengths, spanning from radio waves to $\gamma$-rays. Notably, anomalous behavior is observed in the infrared and $\gamma$-ray spectra, where follows two empirical exponential-type relationships:
\begin{equation}\label{eq1}
S_{q}=A+B\exp\left(-\frac{E}{E_{0}}\right),
\end{equation}
where the parameters $A$ and $B$ for the two domains are $A=2.10\pm 0.92$, $B=0.01\pm 0.023$, and the energy $E_{0}=0.65\pm 0.46$ eV in the range between radio and infrared regime, while $1.91\pm 2.11$, $B=9.04\pm 2.43$, and $E_{0}=0.64\pm 0.18$ eV are the values for the exponential decay in the range between infrared and X-rays. Remarkably, the correlation coefficient stands at 0.51 and 0.75, respectively.

The entropy attains its peak in the radio wave spectrum for energies below the infrared range, starting at 5 GHz. Following this, the entropy experiences a sudden drop at the onset of the infrared range until the X-ray spectrum. Interestingly, the entropy exhibits a sharp increase upon reaching $\gamma$ radiation wavelengths.

The anomalous behavior observed in the entropy-energy relationship of Quasar 3C 273 across different wavelengths, particularly in the infrared and $\gamma$-ray ranges, can be attributed to various physical processes occurring within the quasar. The abrupt increase in entropy in the infrared range followed by an exponential decay until X-rays and then another sudden rise in entropy in the $\gamma$ radiation range may be explained by the interplay of different emission mechanisms and the properties of the Quasar.

The increase in entropy in the infrared range could be linked to processes such as synchrotron radiation, which is common in quasars and can contribute significantly to the emission in this wavelength range \citep{10.1086/422499}. Additionally, the presence of dust-rich environments in quasars, as reported in studies focusing on dust-rich quasars at specific redshifts, could also influence the entropy-energy relationship in the infrared range \citep{10.1088/0004-637x/753/1/33}.

As the entropy then decays exponentially until X-rays, the transition may be associated with changes in the emission mechanisms within the quasar. Differences in X-ray spectra between radio-quiet and radio-loud quasars, as highlighted in studies comparing a large sample of quasars, could play a role in this decay of entropy towards the X-ray range \citep{10.1046/j.1365-8711.2000.03510.x}.

The subsequent rise in entropy when reaching gamma radiation wavelengths might be linked to the extreme luminosity of quasars in this range. Studies on the properties of quasar hosts at the peak of quasar activity and the nuclear-to-host galaxy relation of high-redshift quasars suggest that the energetic processes within the quasar, possibly related to star formation rates and dynamical mass, could contribute to this increase in entropy in the $\gamma$-ray range \citep{10.1088/0004-637x/703/2/1663,10.1086/512847,2023MNRAS.526.4040M}.

In summary, the complex interplay of emission mechanisms, differences in spectra between quasar types, and the energetic processes within the quasar itself likely contribute to the observed anomalous behavior in the entropy-energy relationship of Quasar 3C 273 across different wavelengths.

%%%%%%%%%%%%%%%%%%

Our findings indicate that the energy emanating from Quasar 3C 273 exhibits nonextensive behavior, suggesting that the self-gravitating interactions involved possess long-range characteristics. Additionally, a linear pattern emerges upon analyzing the $q$-entropies of various astrophysical systems across different $q$-entropic indices, as depicted in Fig. \ref{Fig3}. Notably, this figure unveils a linear trend, aligning with the equation:
\begin{equation}\label{eq3}
S_{q}=C+Dq.
\end{equation}
This formulation validates the linear fit, wherein the coefficients are determined as $C=28.12\pm 4.67$ and $D=-12.91\pm 1.98$. Remarkably, the correlation coefficient stands at 0.89, underscoring the robustness of the observed trend.

Figures \ref{Fig2} and \ref{Fig3}, as well as Table \ref{tabresults}, demonstrate a negative value for non-extensive entropy. Generally, nonextensive entropy exhibits a unique characteristic where the total entropy of a system can deviate from the sum of the entropies of its individual components. This deviation can result in negative values of nonextensive entropy, indicating a more organized state with decreased entropy at higher $q$-values, as shown in Fig. \ref{Fig3}. Conversely, lower $q$-values suggest increased entropy and less organization \citep{10.1051/0004-6361:20065151}. Understanding this duality in entropy behavior is crucial for analyzing the physical properties of systems such as quasars.

According to the literature \citep[e.g.,][]{10.1051/0004-6361:20065151,10.1088/0954-3899/36/12/125108}, negative values of nonextensive entropy can indicate the occurrence of phase transitions, providing insights into equilibrium properties, critical phenomena, and the nature of transitions. The correlation between negative nonextensive entropy and phase transitions underscores the importance of incorporating nonextensive statistics in examining the thermodynamic behavior of these physical systems. Specifically, an energy of 200 MeV may be responsible for the anomalous entropy value in $\gamma$-ray emissions.

However, given the two orders of magnitude gap separating it from the last submillimeter value, it remains uncertain whether this mechanism is responsible for the anomalous value near an energy of 0.34 eV. In this interval, the entropy value may be negative or close to zero, which could explain the anomalous value observed in the infrared range. Finally, phase transition mechanisms may occur at specific wavelengths, which is intriguing and could lead to broader discussions about the thermodynamics of quasars.

\section{Summary and future works}
The study of Quasar 3C 273 found that the energy bands predominantly follow Tsallis distributions, with elongated tails indicating a better representation of intensity fluctuations in the light curves. The $q$-Gaussian fitting method was identified as optimal for modeling the light curves, showing high correlation coefficients and significant probability values, outperforming the traditional Gaussian fitting approach. This pattern was consistent across various wavelengths except for the infrared band in 0.34 eV, where a Gaussian profile was more suitable due to the lowest $q$-index value in the dataset.

Moreover, the entropy-energy relationship in Quasar 3C 273 exhibited anomalous behavior, particularly in the infrared and $\gamma$-ray spectra. The entropy peaked in the radio wave spectrum, dropped in the infrared range, decayed exponentially until the X-ray spectrum, and then sharply rose in the $\gamma$ radiation wavelengths. This anomalous behavior was attributed to different emission mechanisms and quasar properties.

The increase in entropy in the infrared range was linked to processes like synchrotron radiation and dust-rich environments within quasars. The subsequent exponential decay until X-rays is associated with changes in emission mechanisms, possibly influenced by differences in X-ray spectra between radio-quiet and radio-loud quasars. The rise in entropy at $\gamma$ radiation wavelengths was connected to the extreme luminosity of quasars in this range, possibly influenced by energetic processes related to star formation rates and dynamical mass within the quasar.

We also verify that the correlation between Tsallis entropy and the $q$-entropic index follows a linear trend. This result suggests that if the wavelength band is more entropic, the value of the associated $q$ entropic index will be lower, and the Quasar 3C 273's nonextensivity will be lower.

%Finally, these findings highlight the complex interplay of various physical processes within Quasar 3C 273, leading to the observed entropy-energy relationships across different wavelengths.
In a forthcoming publication, we aim to investigate non-extensive statistics on various quasars at different redshifts. This endeavor will provide insights into the intricate relationship between quasar energy and the expansion of the Universe. However, we need data with higher redshift values beyond 0.158 from Quasar 3C 273 \citep{1963Natur..197}. With data from the James Webb Space Telescope (JWST), we can scrutinize the spectra of quasars in the early universe. Consequently, this will enable us to map the impacts of cosmic expansion on the energy distribution of quasars.

Another important topic to be addressed in future communication is the galaxy 3C 273 as a source of gravitational waves. \cite{galaxies11050096} analyzed data from the past 60 years of long-term multifrequency monitoring programs of the active galactic nucleus (AGN) 3C 273. The authors proposed a model for estimating the parameters of binary systems near supermassive black holes by analyzing periodic fluctuations in optical and radio bands. For the blazar 3C 273, they determined parameters such as the masses of the companion objects and their orbital properties, including orbital velocities, among others. The authors concluded that AGN 3C 273 could be a massive binary system in an evolutionary stage close to the merger, making it a significant source for detecting gravitational waves. This finding opens an important line of research into the correlation between the system's gravitational wave characteristics and its thermodynamic properties using the formalism of nonextensive statistical mechanics.

\section*{Acknowledgements}
DBdeF acknowledges financial support from the Brazilian agency CNPq-PQ2 (Grant No. 305566/2021-0). The research activities of the STELLAR TEAM of the Federal University of Ceara are supported by continuous grants from the Brazilian agency CNPq.

%% The Appendices part is started with the command \appendix;
%% appendix sections are then done as normal sections

%% If you have bibdatabase file and want bibtex to generate the
%% bibitems, please use
%%
%\bibliographystyle{elsarticle-harv} 
\bibliography{example}

%% else use the following coding to input the bibitems directly in the
%% TeX file.

%%\begin{thebibliography}{00}

%% \bibitem[Author(year)]{label}
%% For example:

%% \bibitem[Aladro et al.(2015)]{Aladro15} Aladro, R., Martín, S., Riquelme, D., et al. 2015, \aas, 579, A101

%%\end{thebibliography}

\end{document}